\documentclass[12pt,preprint]{aastex}
\usepackage{natbib}
\usepackage{rotating}
\usepackage{amssymb,amsmath}

\shorttitle{Ambipolar Diffusion-Mediated Fronts}
\shortauthors{Stone \& Zweibel}

\begin{document}


\title{Ambipolar Diffusion-Mediated Thermal Fronts in the 
Neutral ISM}


\author{Jennifer M. Stone\altaffilmark{1,3} \& 
Ellen G. Zweibel\altaffilmark{1,2,3}}

\altaffiltext{1}{Department of Astronomy, University of Wisconsin--Madison,
    475 N. Charter Street, Madison, WI 53706, USA}
\altaffiltext{2}{Department of Physics, University of Wisconsin--Madison,
    1150 University Avenue, Madison, WI 53706, USA}
\altaffiltext{3}{Center for Magnetic Self-Organization in Laboratory and 
Astrophysical Plasmas}


\begin{abstract}
In a thermally bistable medium, cold, dense gas is separated from warm, 
rareified gas by thin phase transition layers, or fronts, in which heating, 
radiative cooling, thermal conduction, and convection of material are balanced. 
We calculate the steady-state structure of such fronts in the presence of magnetic fields, 
including the processes of ion-neutral drift and ion-neutral frictional heating. 
We find that ambipolar diffusion efficiently transports the magnetic field 
across the fronts, leading to a flat magnetic field strength profile. The 
thermal profiles of such fronts are not significantly different from those of 
unmagnetized fronts. The near uniformity of the magnetic field strength across a 
front is consistent with the flat field strength-gas density relation that is 
observed in diffuse interstellar gas. 
\end{abstract}


\keywords{diffusion---ISM: structure---methods: numerical---MHD}


\section{INTRODUCTION}
The low- and intermediate-temperature parts of the interstellar medium (ISM) 
constitute a thermally bistable medium that results from the balance between 
radiative heating and cooling as well as heating by cosmic rays \citep{FGH69} and
photoelectric heating from PAHs \citep{W95}, the dominant heating 
source. The two stable phases are referred to as the cold neutral medium (CNM), 
having $T_{\rm CNM} \sim 10^{1-2}$ K, and the warm neutral medium (WNM), with 
$T_{\rm WNM} \sim 10^{3-4}$ K. The degree to which magnetic fields are frozen 
into this interstellar gas is parameterized by the magnetic Reynolds number, 
$Re_{\rm M}$, and the ambipolar Reynolds number, $Re_{\rm AD}$ \citep{ZB97}. The magnetic 
Reynolds number is given by the ratio of the Ohmic diffusion time to the 
dynamical time, and for ISM parameters is of order $10^{15} - 10^{21}$. The 
ambipolar Reynolds number, given by the ratio of the ion-neutral drift time to 
the dynamical time, is many orders of magnitude smaller and may approach unity in 
dense molecular gas. Based on these estimates, one would expect that magnetic 
fields should be well coupled to both the ionized part of the gas and to the 
neutrals for all but the most dense or low column density clouds.

Under ideal magnetohydrodynamic (MHD) conditions, such as those indicated above, 
one might expect a strong correlation between magnetic field strength and 
density. If the relationship is expressed as $B \propto \rho^\chi$ and we ignore 
diffusion, for flows directed transverse to the field we have $\chi=1$, whereas 
for field aligned flows $\chi=0$. The median magnetic field strength in the CNM 
has been measured at $B \sim 6 \hspace{0.1cm}  \mu$G \citep{HT05}. If the field 
was frozen in we might expect to detect much smaller field strengths in warmer, 
lower density gas, but instead it is found that the field strength in other ISM 
components is similar to that of the CNM. This was demonstrated by measurements 
of the Zeeman effect over the density range $0.1$ cm$^{-3} < n < 100$ cm$^{-3}$ 
that yielded a flat magnetic field strength-gas density ($B-\rho$) relation 
\citep{TH86}. The most obvious explanation for this relation is that motions are 
aligned with the magnetic field. However, this has been argued against in two 
ways. First, in order for a magnetic field to collimate a flow in this manner it 
must dominate the turbulent energy density, but the field strength is less than 
or equal to equipartition \citep{HZ04}. Second, the accumulation length for the
formation of giant molecular clouds is of order a kiloparsec and may be too large a
scale over which to expect coherent flows \citep{M85}. 
Thus, the flat $B-\rho$ relation may be indicative of magnetic diffusion.   
 
Among the mechanisms that have been proposed to account for the flat $B-\rho$
relation in the diffuse ISM are turbulent ambipolar diffusion (Zweibel 2002;
Heitsch et al. 2004), decorrelation due to MHD waves \citep{PVS03}, and turbulent 
magnetic reconnection (e.g. Santos-Lima et al. 2010). These dynamical studies 
argued that ambipolar diffusion alone was not sufficiently fast to transport 
magnetic flux over the large scales under consideration and so invoked turbulence 
to enhance transport. However, a 1-D two-fluid dynamical study of the thermal 
instability as a formation mechanism for diffuse clouds showed ambipolar
diffusion to efficiently transport magnetic field such that the observed $B-\rho$
relation could be reproduced \citep{IIK07}.
The work presented here complements those findings
but is also a significant departure from that and the other cited examples as we 
shall consider the actual transitions from one phase to another. 
Our approach is advantageous in that we control the diffusive processes and are 
not hampered by numerical diffusion. 
The hydrodynamic structure of CNM/WNM 
transitions has already been presented (Inoue et al. 2006; hereafter IIK06), but 
the effects of a magnetic field have not previously been studied. In the case of 
a magnetic field orthogonal to a transition layer, the field has no effect on 
the structure as it does not exert any force or modify thermal conduction in the 
direction of the temperature gradient, although it has a large effect on
stability \citep{SZ09}. In this work, we consider 
the case of a magnetic field tangential to a transition layer in a simple 1-D 
geometry and include ion-neutral drift as the magnetic diffusion mechanism.    

Our paper is organized as follows: in \S2 we present our numerical method for 
calculating the structure of a phase transition layer for a given initial 
density and magnetic field strength. In \S3 we discuss the effect of ambipolar 
drift heating on the two-phase structure of the neutral ISM. In \S4 we show a 
selection of our ambipolar diffusion-mediated front solutions, and include a 
brief discussion of the flux-freezing approximation. In \S 5 we discuss the 
physical significance of our results, and in \S 6 we summarize our findings.  

\section{METHOD}
\begin{figure}
\epsscale{1.0}
\plotone{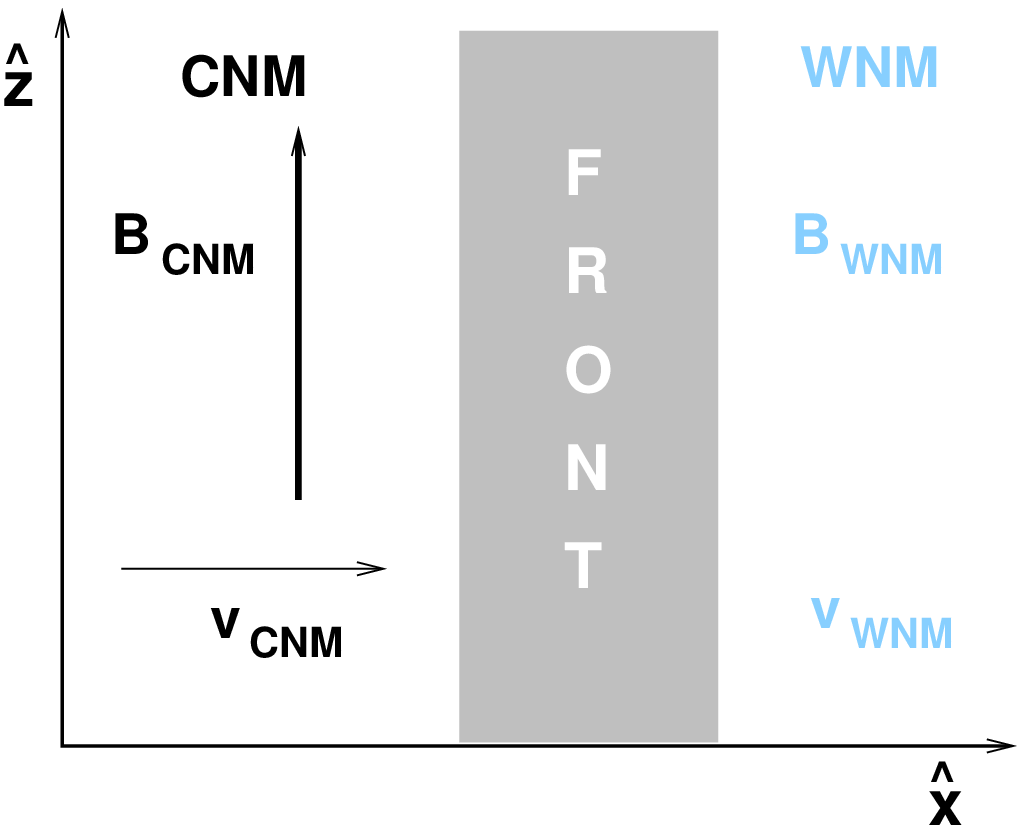}
\caption{\small Geometry of front and magnetic field. We seek a front solution 
connecting the cold neutral medium with the warm neutral medium for the case of 
a uniform magnetic field tangential to the front, $\boldsymbol B_{\rm CNM} = 
B(x)\hat{z}$. The bulk velocity flow is in the $x$-direction. The warm neutral
medium quantities to the right of the front are to be solved for.   
\label{geometry}}
\end{figure}

We consider the scenario of a phase transition layer, or front, separating two 
uniform media of different densities and temperatures in a simple 
one-dimensional geometry with $x$ as the direction of variation. A uniform 
magnetic field is tangential to the front such that 
$\boldsymbol B = B(x)\hat{z}$. The geometry is illustrated in Figure 
\ref{geometry}. We assume a steady-state and ionization equilibrium, and work in 
the reference frame of the front. These assumptions shall be justified in \S 5. 
In order to calculate the physical structure of a front we consider six 
variables: pressure ($p$), density ($\rho$), bulk velocity ($v$), plasma 
velocity ($v_p$), magnetic field strength ($B$), and temperature ($T$), that are 
described by the following five equations, namely the equation of state:
\begin{equation}
\label{eos}       
p = \frac{R \rho T}{\mu} ,
\end{equation}
the continuity equation:
\begin{equation}
\label{cont}
\frac{\partial \rho}{\partial t} + \frac{\partial}{\partial x} \rho v = 0 ,
\end{equation}
the momentum equation:
\begin{equation}
\label{momeq}
\frac{\partial \rho v}{\partial t} + \frac{\partial}{\partial x} 
\Bigg(\rho v^2 + p +\frac{B^2}{8 \pi}\Bigg) = 0 ,
\end{equation}
the induction equation:
\begin{equation}
\frac{\partial B}{\partial t} = - \frac{\partial}{\partial x}(v_p B) ,
\label{induct}
\end{equation}
and the energy equation:
\begin{equation}
\frac{\gamma}{\gamma - 1}\frac{R}{\mu}\rho\frac{dT}{dt} - \frac{dp}{dt}=
\frac{\partial}{\partial x} \kappa \frac{\partial T}{\partial x}  - 
\rho \cal{L} ,
\label{nrg}
\end{equation}
where $\gamma$ is the adiabatic index, R is the molar gas constant, $\mu$ is the
mean molecular weight, $\kappa$ is the thermal conductivity, $\rho \cal{L}$ is 
the cooling function (which includes ambipolar drift heating), and $d/dt \equiv \partial/\partial t + v \hspace{0.1cm} 
\partial/ \partial x$. In the approximation that the plasma and neutral fluids 
are well coupled, and the neutral density dominates, the plasma velocity may be 
written as the sum of the drift velocity, $v_D = v_i - v_n$, and center of mass 
velocity, $v \approx v_n$, such that $v_p \approx v + v_D$, where the drift 
velocity is given by \citep{Shu83}:
\begin{equation}
\label{drift}
\boldsymbol v_D = \frac{ \boldsymbol J \times \boldsymbol B}{c \rho_i \rho_n 
\gamma_{AD}} , 
\end{equation}
with the drag coefficient for collisions between ions and neutrals given by 
$\gamma_{AD} = <\sigma v>_{in}/(m_i + m_n)$ cm$^3$ s$^{-1}$ g$^{-1}$, where 
$<\sigma v>_{in} = 2 \times 10^{-9}$ cm$^3$ s$^{-1}$
\citep{DRD83}. 

Assuming a steady-state (and having already dropped the $\hat{y}$ and $\hat{z}$ 
dimensions), integrating Equations (\ref{cont}), (\ref{momeq}), and (\ref{nrg}) 
with respect to $x$ yields the following conservation laws:
\begin{equation}
j \equiv \rho v , 
\end{equation}
\begin{equation}
M_B \equiv \rho v^2 + p + \frac{B^2}{8 \pi}
\label{MB}
\end{equation}
\begin{equation}
\frac{\gamma}{\gamma - 1}\frac{R}{\mu}j\frac{dT}{dx} - v\frac{dp}{dx}=
\frac{\partial}{\partial x} \kappa \frac{\partial T}{\partial x}  - 
\rho \cal{L} ,
\label{nrg2}
\end{equation}
where $j$ is the mass flux and $M_B$ is the total energy density. To solve 
Equation (\ref{nrg2}) we require expressions describing the evolution of the 
flow speed and magnetic field strength, including the process of ambipolar 
diffusion. An equation for the flow speed is obtained by taking the derivative 
of the total energy density, Equation (\ref{MB}), to obtain:
\begin{equation}
\frac{dv}{dx} = \frac{(\mu v^2 B \frac{dB}{dx} + 4 \pi R j v \frac{dT}{dx})}
{4 \pi j (RT - \mu v^2)} .
\label{vev}
\end{equation}
An equation describing the magnetic field strength is derived by substituting
$\boldsymbol B = B(x)\hat{z}$ into Faraday's law in 1-D and using Equation 
(\ref{induct}) to yield:
\begin{equation}
\frac{\partial B_z}{\partial t} = -c \frac{\partial E_y}{\partial x} 
\Rightarrow c E_y = -(v_p \times B)_y = v_{p x} B_z = \rm constant .
\label{vv}
\end{equation}
Substituting the plasma velocity, with the drift velocity given by Equation 
(\ref{drift}), into Equation (\ref{vv}) we obtain:
\begin{equation}
vB - \frac{B^2}{4 \pi \rho_i \rho_n \gamma_{AD}}\frac{d B}{d x} = 
cE .
\label{magev}
\end{equation}
This is a first-order ODE with one parameter, $cE$. Mathematically, $cE$ can 
take any value since it is a constant of integration. However, we will argue at 
the end of this section that physical considerations of the magnetic field 
strength and ambipolar heating across a front serve to greatly reduce the $cE$ 
parameter space. 

Equations (\ref{nrg2}), (\ref{vev}), and (\ref{magev}), and the definition 
$z \equiv dT/dx$ yield a system of four ODEs for $T$, $B$, and $v$ that apply 
for any functional form of conductivity and cooling function. In the gas states 
studied here, conductivity is dominated by neutral atoms such that 
$\kappa = 2.5 \times 10^3 T^{1/2}$ ergs s$^{-1}$ K$^{-1}$ cm$^{-1}$ \citep{P53}. 
The cooling function is written in full as: 
\begin{equation}
\rho \cal{L} = \rm n [n \Lambda - (\Gamma_{PAH} + \Gamma_{AD})] .
\label{TE}
\end{equation}  
We take the simple functional forms used by IIK06 for Ly$\alpha$ and [C II]
radiative cooling such that:
\begin{equation}
\Lambda = 7.3 \times 10^{-21} \rm{exp} \Bigg(\frac{-118400 \hspace{0.1cm} K}
{T+1500 \hspace{0.1cm} K}\Bigg)+
7.9 \times 10^{-27} \rm{exp} \Bigg(\frac{-92 \hspace{0.1cm} K}{T}\Bigg) 
\rm ergs \hspace{0.2cm} s^{-1} cm^{-3} ,
\label{lambdacool}
\end{equation}
and for photoelectric heating: 
\begin{equation}
\Gamma_{\rm PAH} = 2 \times 10^{-26} \rm ergs \hspace{0.2cm} s^{-1} .
\label{gammapah}
\end{equation}
Heating by ion-neutral friction is represented by $\Gamma_{AD}$:
\begin{equation}
\label{adheat}
n \Gamma_{\rm AD} = \rho_i \rho_n \gamma_{\rm AD} v_D^2 =
\frac{1}{\rho_i \rho_n \gamma_{\rm AD}} \Bigg(\frac{B}{4 \pi} 
\frac{d B}{d x}\Bigg)^2 \rm ergs \hspace{0.2cm} s^{-1} cm^{-3}
\end{equation}  
(Scalo 1977, Padoan et al. 2000), where the density of neutrals is given by 
$\rho_n = \mu_n m_H n$, and the density of ions by $\rho_i = \mu_i m_p n_e$. We 
compute the ionization fraction using:
\begin{eqnarray} 
\frac{n_e}{n_H} & = & \Bigg(1.19 \times 10^{-4} - 1.36 \times 10^{-8} 
\frac{T^{0.845}}{n_H}\Bigg) + \nonumber \\
& &
\Bigg(1.42 \times 10^{-8} + 2.72 \times 10^{-8} \frac{T^{0.845}}{n_H}
+ 1.85 \times 10^{-16} \frac{T^{1.69}}{n_H^2}\Bigg)^{1/2}
\end{eqnarray} 
\citep{FZS88}.

In solving this system we choose initial values for the density and magnetic 
field strength and impose the following boundary conditions:  
\begin{equation}
\label{Tbc}
T(x=x_1) = T_1, \hspace{1in} T(x=x_2) = T_2,
\end{equation}
\begin{equation}
\label{dTbc}
\frac{dT}{dx} \Bigg |_{x_1,x_2} = 0 ,
\end{equation}
where $x_1$ and $x_2$ represent the left- and right-hand boundaries, 
respectively, and $T_1$ and $T_2$ satisfy thermal equilibrium at these 
boundaries. $T_1$ is found by solving Equation (\ref{TE}) for the chosen initial 
value of the density at $x_1$, and $T_2$ is the temperature obtained by 
integrating as far as $x_2$, where the size of domain is chosen such that $T_2$ 
will also satisfy thermal equilibrium. For our third and fourth conditions, 
given by Equation (\ref{dTbc}), we impose zero temperature gradient at both 
boundaries. Finally, we set the value of the initial magnetic field strength 
gradient, $|dB/dx|_{x_1}$, as this controls the amount of ambipolar heating in a 
given front model. As we will show in \S4, the choice of the initial field 
strength gradient affects the structure of the front. Given that we set five 
boundary values but have a system of only four ODEs, we thus set up an 
eigenvalue problem in which the mass flux, $j$, is the parameter to be adjusted 
to find a self-consistent solution. 

The numerical method we employ is that of shooting, in which the integration is 
performed with an initial guess for $j$, the resulting boundary values compared 
to the desired conditions, and $j$ adjusted accordingly so that the integration 
can be repeated as necessary until the right-hand boundary conditions are 
satisfied to within some chosen tolerance. We find that the degree to which 
thermal equilibrium is satisfied at the right-hand boundary depends on the size 
of the domain, which should be adjusted to achieve optimum results. For cases in 
which ambipolar drift heating does not dominate it is possible to satisfy 
thermal equilibrium to better than one part in $10^5$. We use a 5th-order 
adaptive Runge-Kutta scheme \citep{Numrec} and adjust the eigenvalue according to the secant method. 
When appropriate bounds are chosen our method converges to a solution quickly, 
requiring of order ten iterations. Note that we always integrate from the cold 
medium to the warmer one.  

We close this section with a brief discussion of the initial magnetic field
strength gradient boundary condition and the parameter $cE$. In setting up our 
initial conditions, instead of choosing the value of $cE$ directly we instead 
set the initial value of the magnetic field strength gradient, $|dB/dx|_{x_1}$. 
This implies the value of $cE$, which is kept constant across the domain, as we 
may evaluate it by substituting our initial conditions into Equation 
(\ref{magev}). Note that the value of $cE$ will change with each iteration of 
the shooting method because it depends on the bulk velocity, which is adjusted 
according to the secant method. For all density and magnetic field strength 
initial conditions there is some minimum value of $cE$ below which the magnetic
field strength gradient is positive throughout the domain. The outcome for 
choosing an initial field gradient that yields a value of $cE$ below this 
minimum would be a larger magnetic field strength in the warm medium than in the 
colder one. However, if one imagines an evaporating cool cloud with the
assumption of frozen-in magnetic field lines, this does not seem like a 
physically reasonable scenario as the field lines will become further apart as 
the cloud expands. Furthermore, if the value of cE is too large, ambipolar drift 
heating may dominate over photoelectric heating making it increasingly difficult to 
satisfy thermal equilibrium at the far boundary, implying that a front can no 
longer exist. We demonstrate quantitatively the effects of $|dB/dx|_{x_1}$, and
hence $cE$, in \S 4, but do not refer to $cE$ explicitly in the rest of the 
paper.  

\section{EFFECTS OF AMBIPOLAR DRIFT HEATING ON TWO-PHASE STRUCTURE}
\begin{figure}
\epsscale{1.0}
\plotone{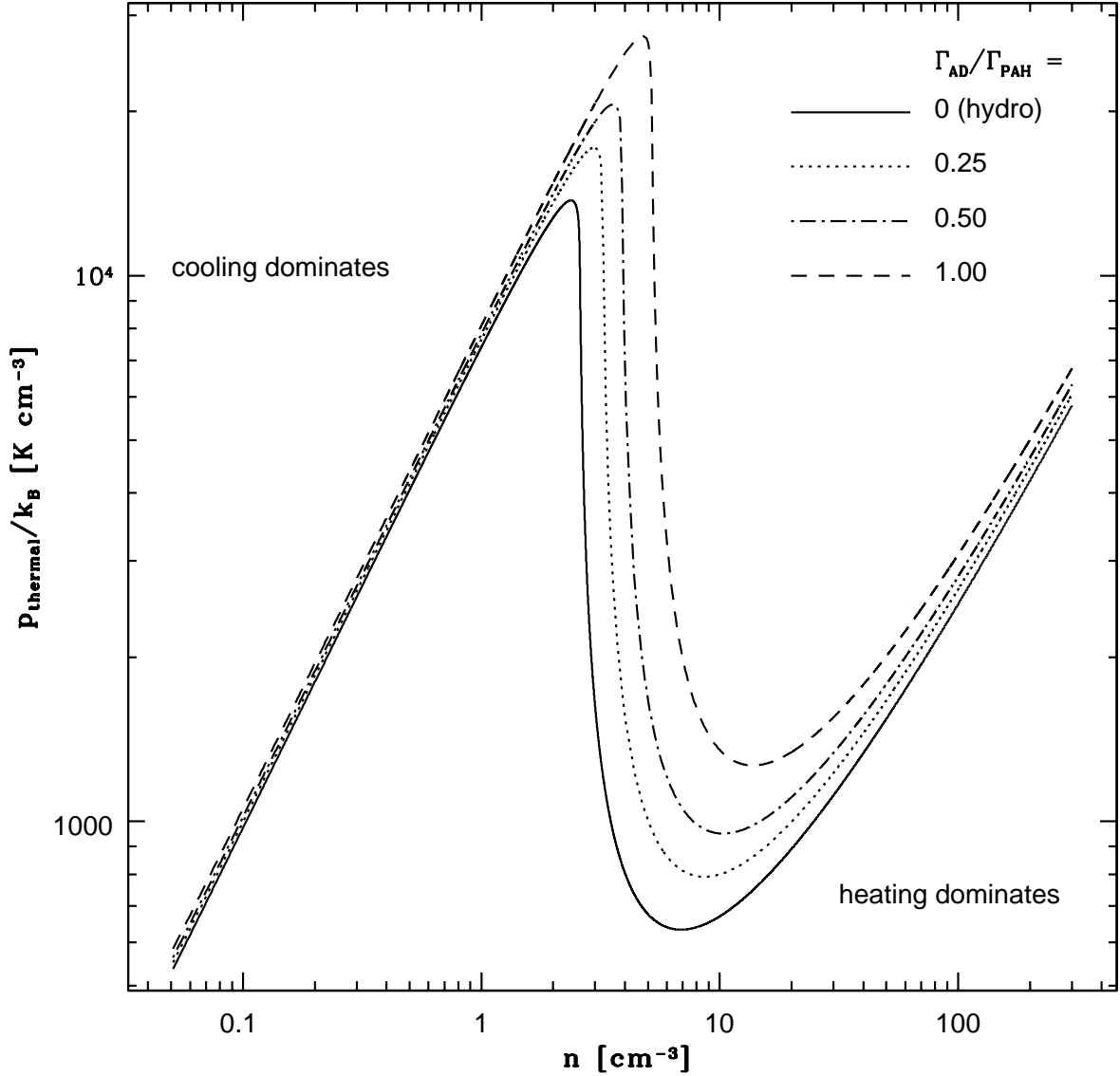}
\caption{\small Thermal equilibrium state of the cooling function, $\rho
\cal{L} \rm (n,T) =0$, in the thermal pressure-number density plane. The solid 
line shows the hydrodynamic case in which there is no ambipolar drift heating. 
The other curves show the effect of increasing the ambipolar drift heating rate 
at a fixed magnetic field strength of $3 \hspace{0.1cm} \mu$G. In the area above 
the curves $\rho \cal{L} \rm > 0$ so cooling dominates, while below the curves 
$\rho \cal{L} \rm < 0$ so heating dominates. Increasing the ambipolar drift 
heating rate increases the pressure at which two phases can co-exist. Note that 
the ambipolar drift heating rate is actually a function of density and magnetic 
field strength, as given by Equation (\ref{adheat}), but for the purposes of 
this plot we set it to be some constant fraction of the photoelectric heating rate. 
\label{te}}
\end{figure}

The two neutral phases of the ISM are enabled by the balance of radiative
cooling and heating by, in this work, photoelectric heating and ion-neutral friction. We
present the equilibrium state of the cooling function, $\rho \cal{L} \rm (n, T)
= 0$, in Figure \ref{te}, with $\rho \cal{L} \rm$ given by Equation (\ref{TE}).
The solid line shows the case in which there is no ambipolar drift heating, for 
which IIK06 report that a two-phase structure is possible for $10^{2.8}$ K
cm$^{-3} < p/k_B < 10^{4.1}$ K cm$^{-3}$. The other lines illustrate the effects
of increasing the ambipolar drift heating rate at a fixed magnetic field 
strength of $3 \hspace{0.1cm} \mu$G. Although we have already shown 
$\Gamma_{AD}$ to be a function of the density and field strength, for the 
purposes of this plot we have set it to be a constant fraction of the photoelectric
heating rate, $\Gamma_{\rm PAH}$. Increasing the total heating rate serves to 
increase the pressure at which two phases can co-exist: the minimum pressure at 
which the cold phase can exist, and the maximum pressure at which the warmer 
phase can exist, both increase. In fact, the pressure range over which two 
phases can exist becomes larger as the total heating is increased. For example, 
for the $\Gamma_{\rm AD}/\Gamma_{\rm PAH} = 0.50$ case plotted in Figure \ref{te}, 
two-phase structure is possible for $10^{3.0}$ K cm$^{-3} < p/k_B < 10^{4.3}$ K 
cm$^{-3}$, and for the $\Gamma_{\rm AD}/\Gamma_{\rm PAH} = 1.00$ case the pressure range 
is $10^{3.1}$ K cm$^{-3} < p/k_B < 10^{4.4}$ K cm$^{-3}$. We can understand the 
shift towards lower densities as follows: increasing the heating increases the 
temperature, so it must decrease the density. The upshift of the equilibrium to 
higher pressures also reflects the increased heating.

\section{FRONT SOLUTIONS}
The characteristics of a front are determined by its thermal pressure. There 
exists a ``saturation pressure'' at which heating and cooling are balanced 
within a front (Zel'dovich \& Pikel'ner 1969, Penston \& Brown 1970). If 
$\Lambda$ and $\Gamma$ can be written as functions of pressure and temperature 
(where $\Gamma$ is the total heating rate), this pressure may be calculated by 
solving the integral 
\citep{IIK06}:
\begin{equation}
\int_{T_1}^{T_2} \kappa \rho \cal{L} \rm dT = 
\int_{T_1}^{T_2} \kappa n (n \Lambda - \Gamma) dT = 0 
\label{psat}
\end{equation}
and substituting for $n$ using the equation of state, Equation (\ref{eos}). 
For the hydrodynamic case ($\Gamma_{AD} = 0$) IIK06 obtain $p_{sat}/k_B = 2612$ K 
cm$^{-3}$ (where $k_B$ is the Boltzmann constant), which, by solving Equation (\ref{TE}), implies an initial density of 
$n = 106.08$ cm$^{-3}$ and hence an initial temperature of $T = 24.63$ K. If the 
thermal pressure exceeds this value of $p_{sat}$ a fluid element passing through 
the front experiences net cooling, so we have a condensation front. If instead 
the thermal pressure is less than the saturation value a fluid element 
experiences net heating, so we have an evaporation front. 

In this section we demonstrate the effect of ambipolar diffusion on the 
saturation pressure and present our ambipolar diffusion-mediated front 
solutions. We also argue that the flux-freezing approximation is not accurate 
for steady-state thermal fronts.

\subsection{Effects of Ambipolar Drift Heating on Saturation Pressure}  
\begin{figure}
\epsscale{1.0}
\plotone{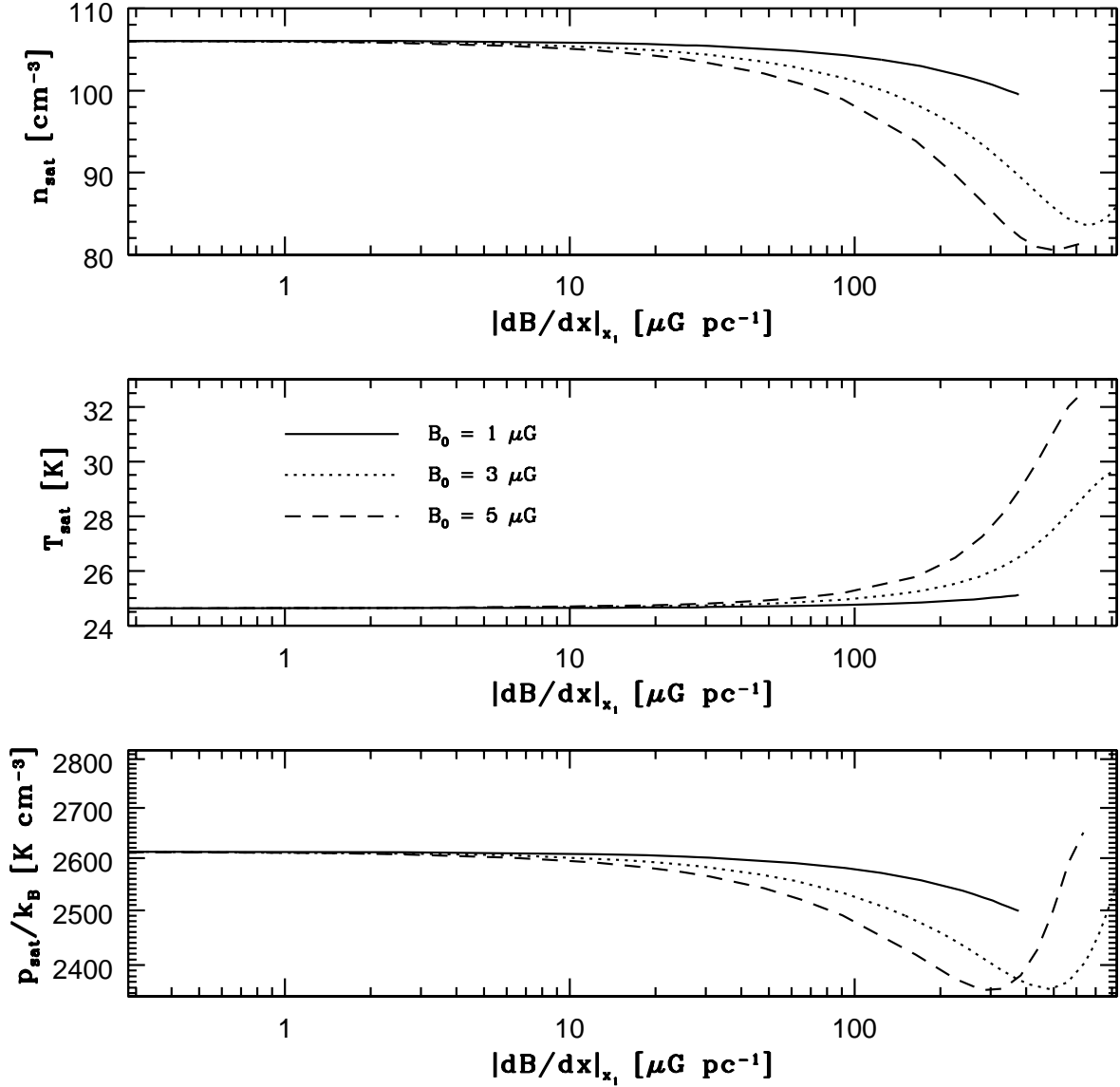}
\caption{\small Saturation number density, temperature, and pressure (where
$p_{sat}/k_B = n_{sat} T_{sat}$) as a function of the initial magnetic field 
strength gradient for initial field strengths of $1$, $3$, and $5 \hspace{0.1cm} 
\mu$G. The range of $|dB/dx|_{x_1}$ has been chosen to show the inflection point 
of the saturation pressure. The $1 \hspace{0.1cm} \mu$G curve has not been 
extended further because when $|dB/dx|_{x_1}$ is too large such a front cannot 
connect to a thermal equilibrium phase before the magnetic field strength 
becomes negative. The $3$ and $5 \hspace{0.1cm} \mu$G curves can be extended to
larger saturation pressures than shown here, until their magnetic profiles also
become negative.
}
\label{Bpsat}
\end{figure}

The saturation pressure is altered in the presence of a magnetic field due to
ambipolar drift heating. The integral given by Equation (\ref{psat}) cannot be 
solved analytically when $\Gamma_{\rm AD}$ is non-zero, so instead we use our 
shooting method, as discussed in \S2, to find the initial density that yields a 
static solution as a function of the initial magnetic field strength gradient. 
The results for initial magnetic field strengths of $1$, $3$, and $5 
\hspace{0.1cm} \mu$G are shown in Figure \ref{Bpsat}. Note that the field 
gradients are actually negative, as we anticipate the magnetic field strength to 
decrease with increasing temperature, and we refer to the absolute magnitude of 
the quantity, which we give in units of $\mu$G pc$^{-1}$. 

As $|dB/dx|_{x_1}$ is 
increased the saturation density and pressure for all magnetic field strengths 
initially decrease until a sufficiently large value of $|dB/dx|_{x_1}$ is 
reached, after which the density and pressure both increase. Therefore it is 
possible to have two different fronts at the same saturation pressure. This
nonmonotonic behavior may be understood by solving Equation (\ref{psat}) for the saturation
pressure using the equation of state, Equation (\ref{eos}), to obtain:
\begin{equation}
\frac{p_{sat}}{k_B} = \frac{\Gamma \int_{T_1}^{T_2} \frac{\kappa}{T}  dT}
{\int_{T_1}^{T_2} \frac{\kappa \Lambda}{T^2}  dT} .
\label{psat2}
\end{equation}
This shows that increasing the total heating rate, $\Gamma$, 
tends to increase the saturation pressure. But increased heating also tends to
drive up the temperature, which for CNM temperatures greatly increases the
cooling rate, $\Lambda$, and according to Equation (\ref{psat2}) this decreases the
saturation pressure. For example, Figure \ref{Bpsat} shows that if $B_0 = 5 
\hspace{0.1cm} \mu$G and $|dB/dx|_{x_1} = 300 \hspace{0.1cm} \mu$G pc$^{-1}$ the CNM temperature
is increased from $24$ to $28$ K. According to Equation (\ref{lambdacool}), this
results in a greater than $70 \%$ increase in the cooling rate, $\Lambda$.
Such a large increase
in cooling requires a lower density and a lower saturation pressure for
equilibrium to be maintained. This effect dominates as long as the heating and
cooling rates, $\Gamma$ and $\Lambda$, are not too large and is the reason for 
the dip in the saturation pressure seen in Figure \ref{Bpsat}. 
An inflection point is not observed in the saturation pressure in the $1 
\hspace{0.1cm} \mu$G case, the reason being that at higher values of 
$|dB/dx|_{x_1}$ (and hence larger ambipolar heating rates) the magnetic field 
profile is so steep that a thermal equilibrium phase cannot be reached before 
the magnetic field strength becomes negative. In such instances the temperature 
on the cold side is still well within the range of CNM values, so it is not the 
medium being overheated that prohibits physical front solutions. 

\begin{figure}
\epsscale{1.0}
\plotone{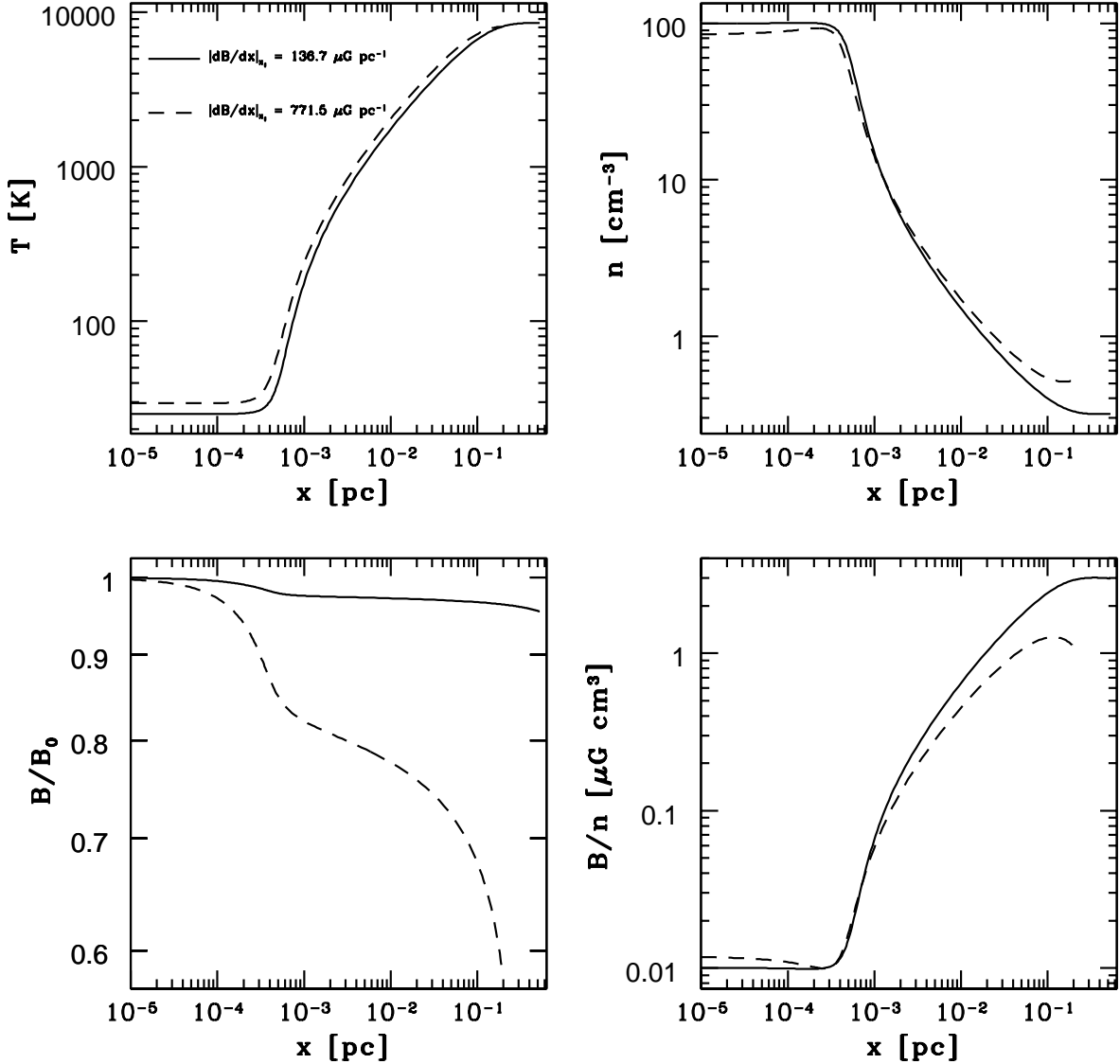}
\caption{\small Two static fronts with $p_{\rm thermal}/k_B = 2500$ K cm$^{-3}$ and an 
initial magnetic field strength of $3 \hspace{0.1cm} \mu$G, but different 
initial magnetic field strength gradients. Only the phase transition is shown, 
so the reader may find it helpful to picture the cold phase occupying the region 
$x < 0$, and the warmer medium filling the region beyond the end of the 
transition. The dashed line shows the front subject to a higher ambipolar 
diffusion heating rate, which connects a cold medium, with $n_{\rm CNM} = 84.66$ 
cm$^{-3}$ and $T_{\rm CNM} = 29.53$ K, with a warmer medium with $n_{\rm WNM} = 
0.52$ cm$^{-3}$ and $T_{\rm WNM} = 8063$ K. The front with the lower ambipolar 
drift heating rate connects a cold phase having $n_{\rm CNM} = 95.51$ cm$^{-3}$ 
and $T_{\rm CNM} = 25.12$ K  with a warm phase having $n_{\rm WNM} = 0.32$ 
cm$^{-3}$ and $T_{\rm WNM} = 8533$ K. The top panels show the temperature and 
density profiles and the lower panels show the magnetic field strength profiles 
and the ratio of the field strength to the density for both models.
\label{doublestatic}}
\end{figure}

We present example saturation fronts having a thermal pressure of $p_{th}/k_B = 
2500$ K cm$^{-3}$ and an initial field strength of $3 \hspace{0.1cm} \mu$G, but 
with different initial values of $|dB/dx|_{x_1}$, in Figure \ref{doublestatic}. 
Note that the values of $|dB/dx|_{x_1}$ given represent the largest gradients at 
any point throughout the front. The magnetic field strength gradients of all the 
fronts we present quickly relax to become much smaller than the initial values 
that we impose. The front having the larger value of $|dB/dx|_{x_1}$ has a 
higher ambipolar drift heating rate and connects a lower density, higher 
temperature CNM with a higher density, cooler WNM than the static front with the 
lower heating rate. The front with the lower ambipolar drift heating rate is the 
most diffusive, which is illustrated by its flatter magnetic field strength 
profile. This is also indicated by the ratio of the field strength to the number 
density, which shows a larger variation across the domain than the same quantity 
for the static front with the higher heating rate.  

\subsection{Ambipolar Diffusion-Mediated Front Solutions}
\begin{figure}
\epsscale{1.0}
\plotone{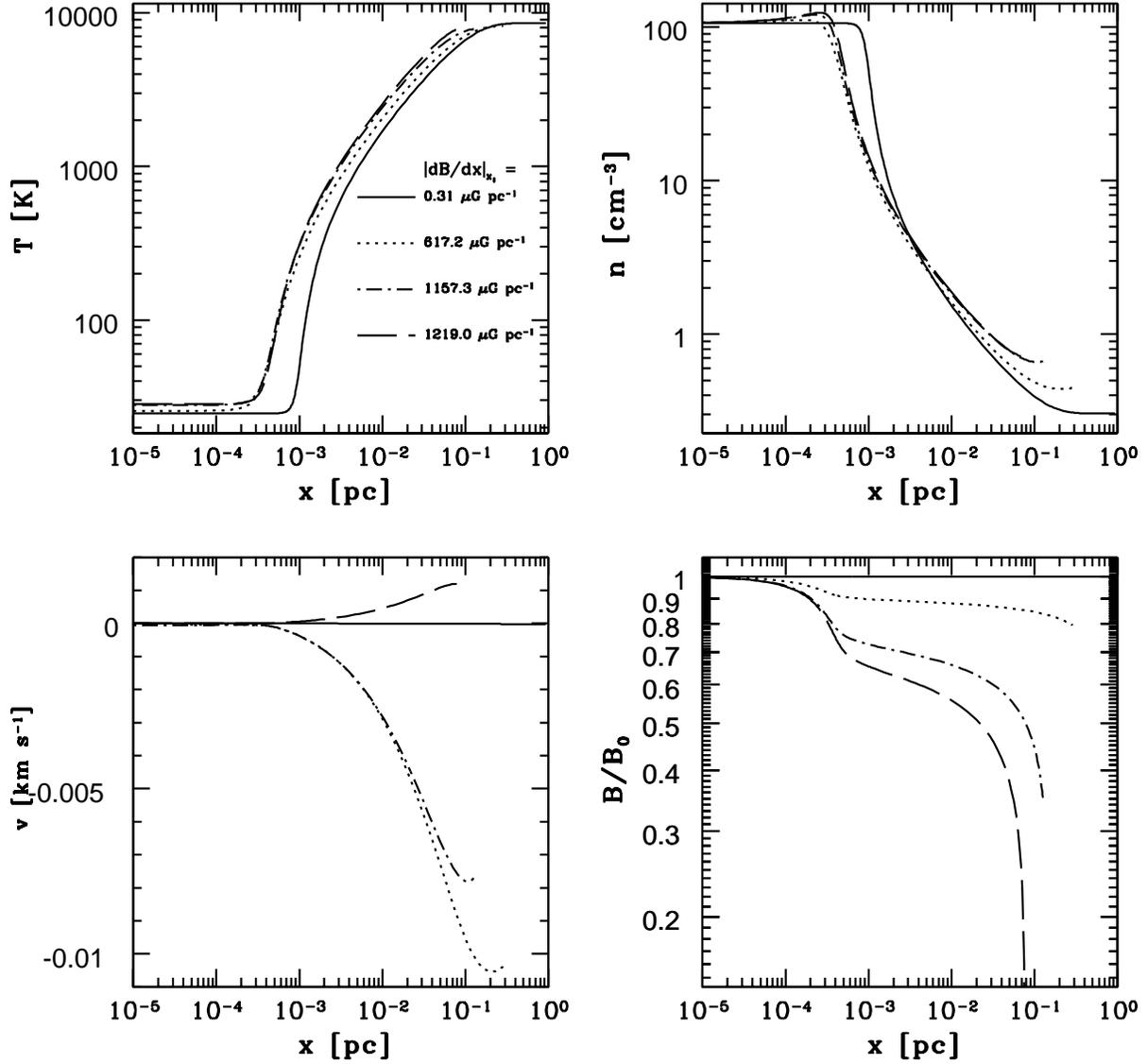}
\caption{\small Profiles of fronts having an initial density of $n = 106.08$ 
cm$^{-3}$ and an initial magnetic field strength of $B_0 = 3 \hspace{0.1cm} 
\mu G$ for various initial magnetic field strength gradients. The top panels 
show the temperature and density profiles and the lower panels show the bulk 
velocity 
and magnetic field strength profiles of the fronts. The different line styles 
represent different values of $|dB/dx|_{x_1}$ (note that these same line styles 
are used in Figures \ref{driftfig} and \ref{heat}).
\label{prof1}}
\end{figure}

As stated at the beginning of \S 4, in the hydrodynamic case a static front has 
a thermal pressure of $p_{sat}/k_B = 2612$ K cm$^{-3}$, which corresponds to an 
initial density and temperature of $n = 106.08$ cm$^{-3}$ and $T = 24.63$ K, 
respectively. To demonstrate the effects of ambipolar diffusion we present 
several front models having this same initial density and an initial magnetic 
field strength of $3 \hspace{0.1cm} \mathrm{\mu G}$ in Figure \ref{prof1}. The 
initial temperature changes according to the ambipolar drift heating rate, so is 
not the same as in the hydrodynamic case. The different models correspond to 
various initial magnetic field strength gradients, $|dB/dx|_{x_1}$, where a 
larger gradient corresponds to increased heating. The properties of the phases 
connected by these fronts are listed in Table 1. 

The overall shapes of the temperature profiles, shown in the top left panel of
Figure \ref{prof1}, are fairly similar with the main differences being the 
temperature gradients on small scales and the final temperatures of the warm 
phases becoming lower as $|dB/dx|_{x_1}$ is increased. The main effect of 
increasing $|dB/dx|_{x_1}$ is that the size of the integration domain required 
to reach thermal equilibrium at the right-hand boundary becomes smaller, due to 
the increased ambipolar drift heating. In fact, the lowest $|dB/dx|_{x_1}$ 
profile shown here is very similar to the hydrodynamic solution of IIK06. The 
top right panel of Figure \ref{prof1} shows that the density varies by more than 
two orders of magnitude across the front for all heating rates. As 
$|dB/dx|_{x_1}$ is increased the density of the warm phase at the far boundary 
increases and hence the temperature decreases. 

For insight into the actual nature of fronts, one may begin by looking at the 
bulk velocity profiles, shown in the bottom left panel of Figure \ref{prof1}. The 
effect of the initial magnetic field strength gradient on the velocity profile 
of a front is not straightforward. For the lowest initial $|dB/dx|_{x_1}$ case 
shown the velocity profile is flat and close to zero, as should be the case for 
a static front. As $|dB/dx|_{x_1}$ is increased the velocity at first becomes 
larger and negative. This is because the saturation pressure is altered from the 
original hydrodynamic value of $p_{sat}/k_B = 2612$ K cm$^{-3}$, as we discussed 
in \S4.1. The models shown here with negative velocity profiles are actually 
condensation fronts. However, as $|dB/dx|_{x_1}$ is further increased there 
comes a point when the velocity no longer becomes increasingly negative, and 
instead begins to increase. Eventually the front transitions from being a 
condensation front to an evaporation front, which is illustrated by the positive 
velocity profile of the largest $|dB/dx|_{x_1}$ model shown in Figure 
\ref{prof1}. This is expected because of the nonmonotonic behavior of the
saturation pressure as the ambipolar heating rate is increased (see Figure 
\ref{Bpsat}). 

The magnetic field strength profile of the lowest value $|dB/dx|_{x_1}$ model, 
given in the lower right panel of Figure \ref{prof1}, is extremely flat. As 
$|dB/dx|_{x_1}$ is increased the field strength decreases across the domain in 
an almost linear fashion; for sufficiently large values the profile becomes 
nonlinear. Given that the change in density across a front is much more dramatic
than that of the magnetic field strength, the ratio of the magnetic field 
strength to the number density of neutrals, $B/n$, changes markedly throughout 
the transition layer.  

\begin{figure}
\epsscale{1.0}
\plotone{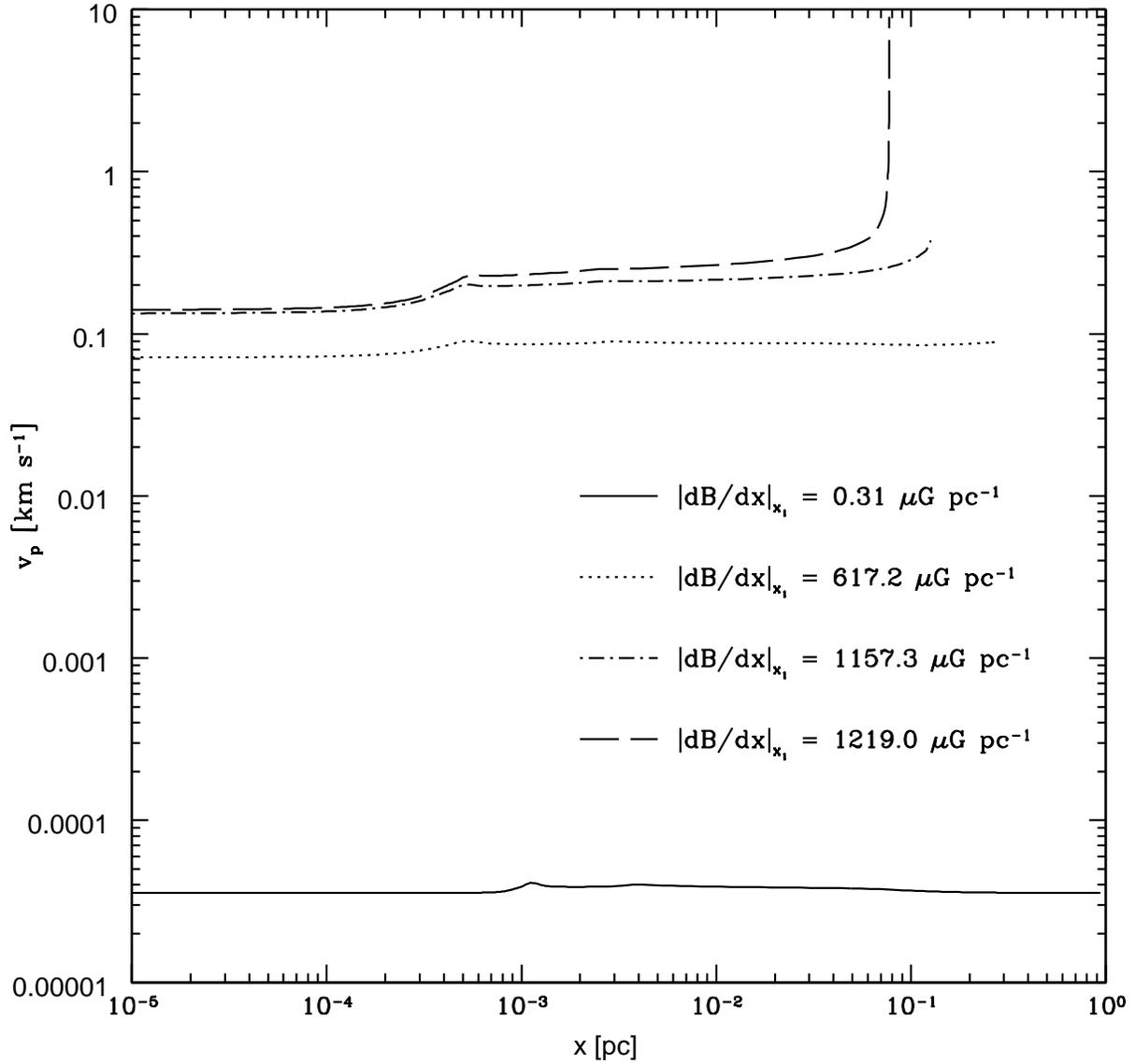}
\caption{\small Plasma velocity ($v_p \approx v + v_D$) profiles of fronts having an initial density of $n = 106.08$ 
cm$^{-3}$ and an initial magnetic field strength of $B_0 = 3 \hspace{0.1cm} 
\mu G$ for various initial magnetic field strength gradients.
\label{driftfig}}
\end{figure}

In Figure \ref{driftfig} we plot the plasma velocity profile, given by $v_p
\approx v + v_D$ and Equation (\ref{drift}), of each of the front models of 
Figure \ref{prof1}. The shapes of the profiles are governed by the behavior of 
the magnetic field strength. The plasma velocity is
almost constant across the lowest $|dB/dx|_{x_1}$ model since the magnetic field
strength profile is close to flat, whereas the larger $|dB/dx|_{x_1}$ models show more 
variation in their plasma velocity profiles due to the presence of significant 
gradients in the magnetic field. In all cases the drift velocity is positive and 
larger than the bulk velocity of the front, such that the plasma velocity is also
positive.

\begin{figure}
\epsscale{1.0}
\plotone{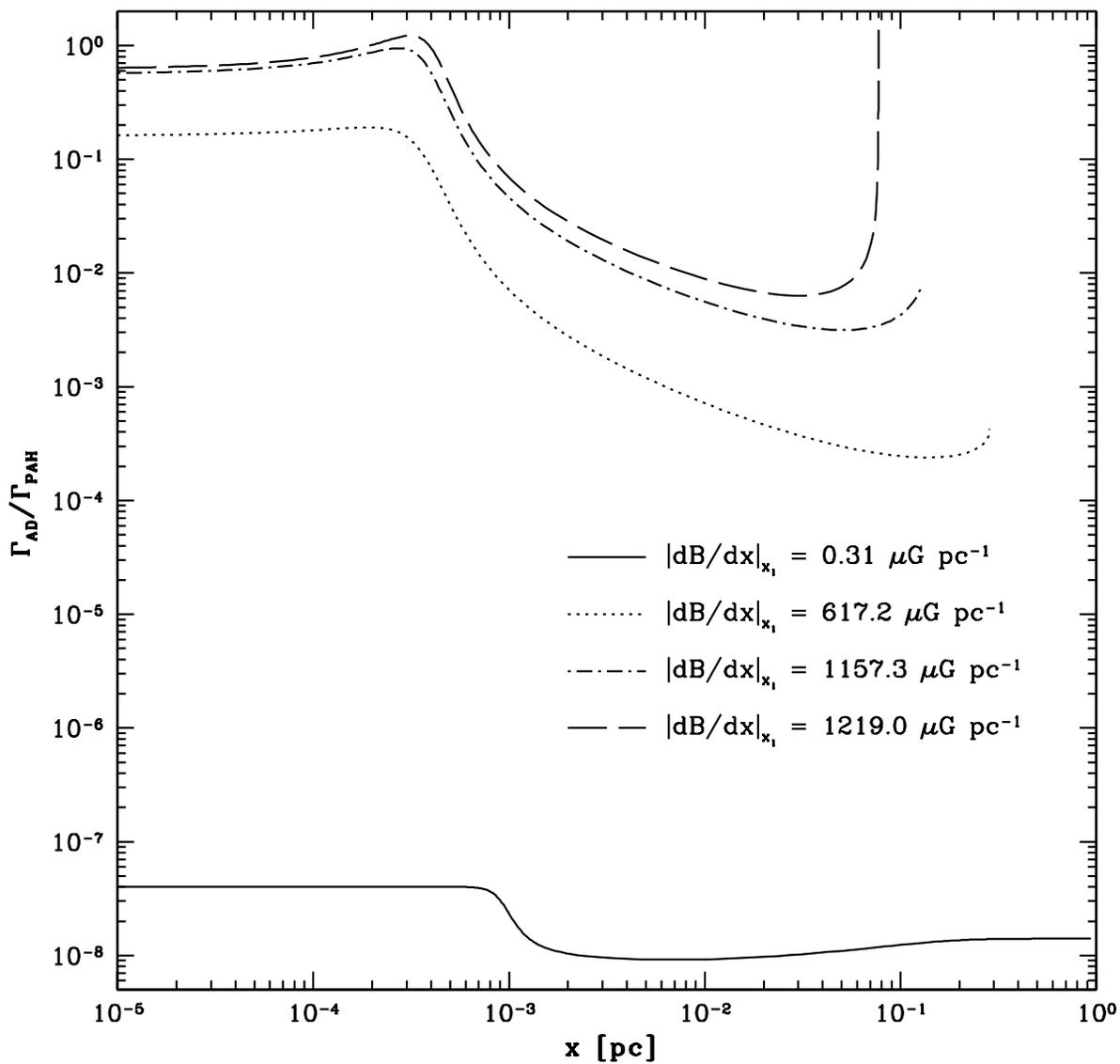}
\caption{\small Ambipolar drift heating rates normalized by photoelectric 
heating rate across fronts having an initial density of $n = 106.08$ cm$^{-3}$ 
and an initial magnetic field strength of $B_0 = 3 \hspace{0.1cm} \mu G$, for 
various initial field strength gradients.  
\label{heat}}
\end{figure}

In Figure \ref{heat} we compare the ion-neutral drift heating rate, given by
Equation (\ref{adheat}), to that of photoelectric heating, given by Equation
(\ref{gammapah}), for each of the front models of Figure \ref{prof1}. The lowest 
$|dB/dx|_{x_1}$ model has a much smaller ambipolar drift heating rate than 
photoelectric heating rate which is why the structure of that front is barely 
different from the hydrodynamic case. The three larger $|dB/dx|_{x_1}$ fronts 
have larger ambipolar heating rates that are comparable to the photoelectric 
heating rate. These fronts depart more noticeably from the hydrodynamic 
solution and are less diffusive.    

\begin{figure}
\epsscale{1.0}
\plotone{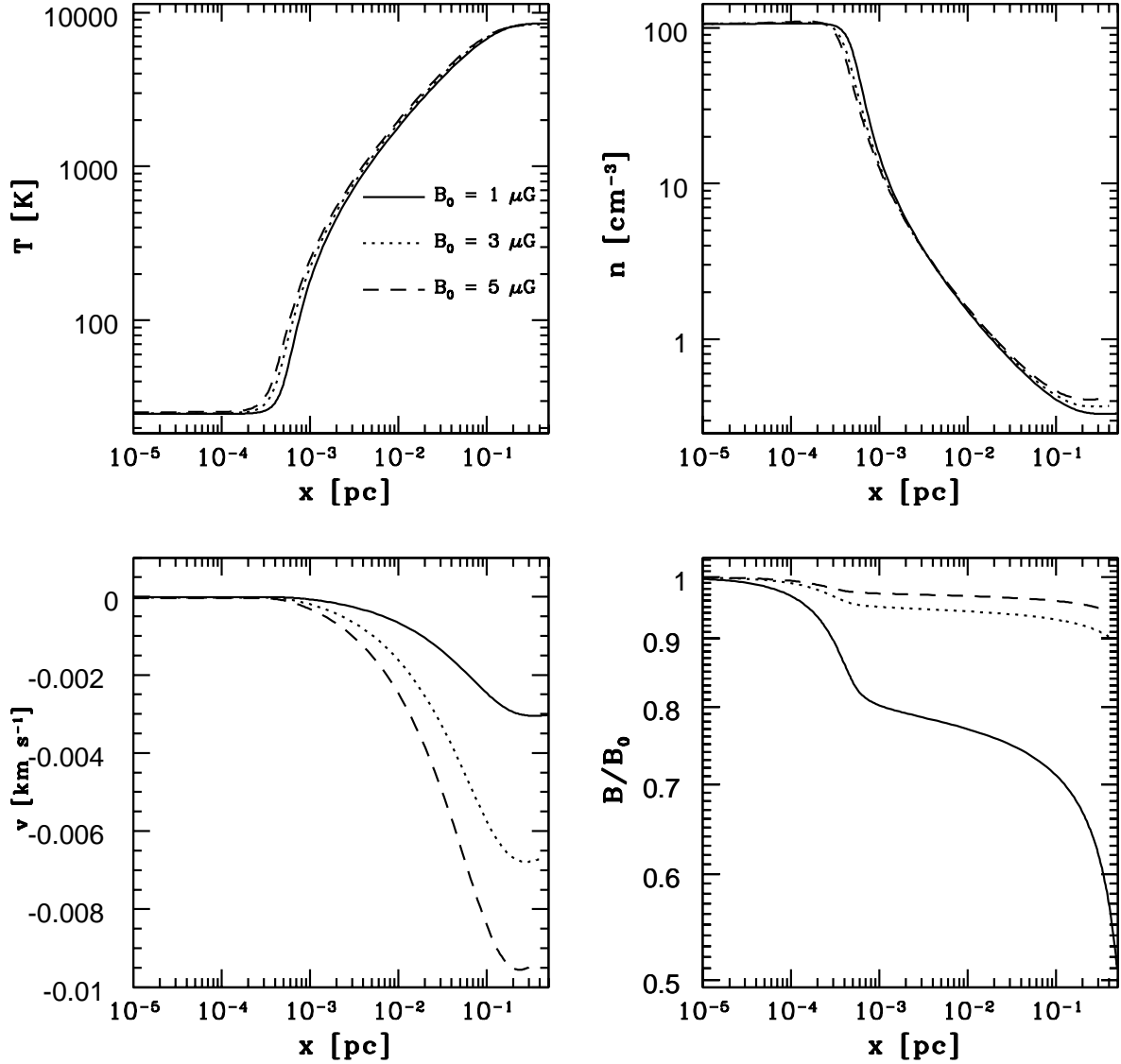}
\caption{\small Temperature, density, bulk velocity, and magnetic field strength
profiles of fronts having an initial density of $n = 106.08$ cm$^{-3}$ at 
various magnetic field strengths for fixed $|dB/dx|_{x_1} = 308.6 \hspace{0.1cm} 
\mu$G pc$^{-1}$. The same line styles are also employed in Figures
\ref{Bdriftfig} and \ref{Bheat}. }
\label{Bfig}
\end{figure}

We also investigate the effect of magnetic field strength on front profiles at 
fixed initial $|dB/dx|_{x_1}$. Figure \ref{Bfig} shows a variety of front 
characteristics for inital field strengths of $1$, $3$, and $5 \hspace{0.1cm} 
\mu$G and $|dB/dx|_{x_1} = 308.6 \hspace{0.1cm} \mu$G pc$^{-1}$. The temperature 
profiles are very similar, with the effect of increasing the field strength 
being larger temperature gradients at small scales and thinner fronts. The effect on 
the density 
profile is that the higher magnetic field strength fronts connect warm phases with higher 
densities. The velocity profiles are slightly negative, which implies that these 
are actually condensation fronts, and the departure from a static solution seems 
to increase with increasing field strength. Higher field strength solutions have 
flatter magnetic profiles because the efficiency of ambipolar diffusion 
increases with magnetic field strength. 

\begin{figure}
\epsscale{1.0}
\plotone{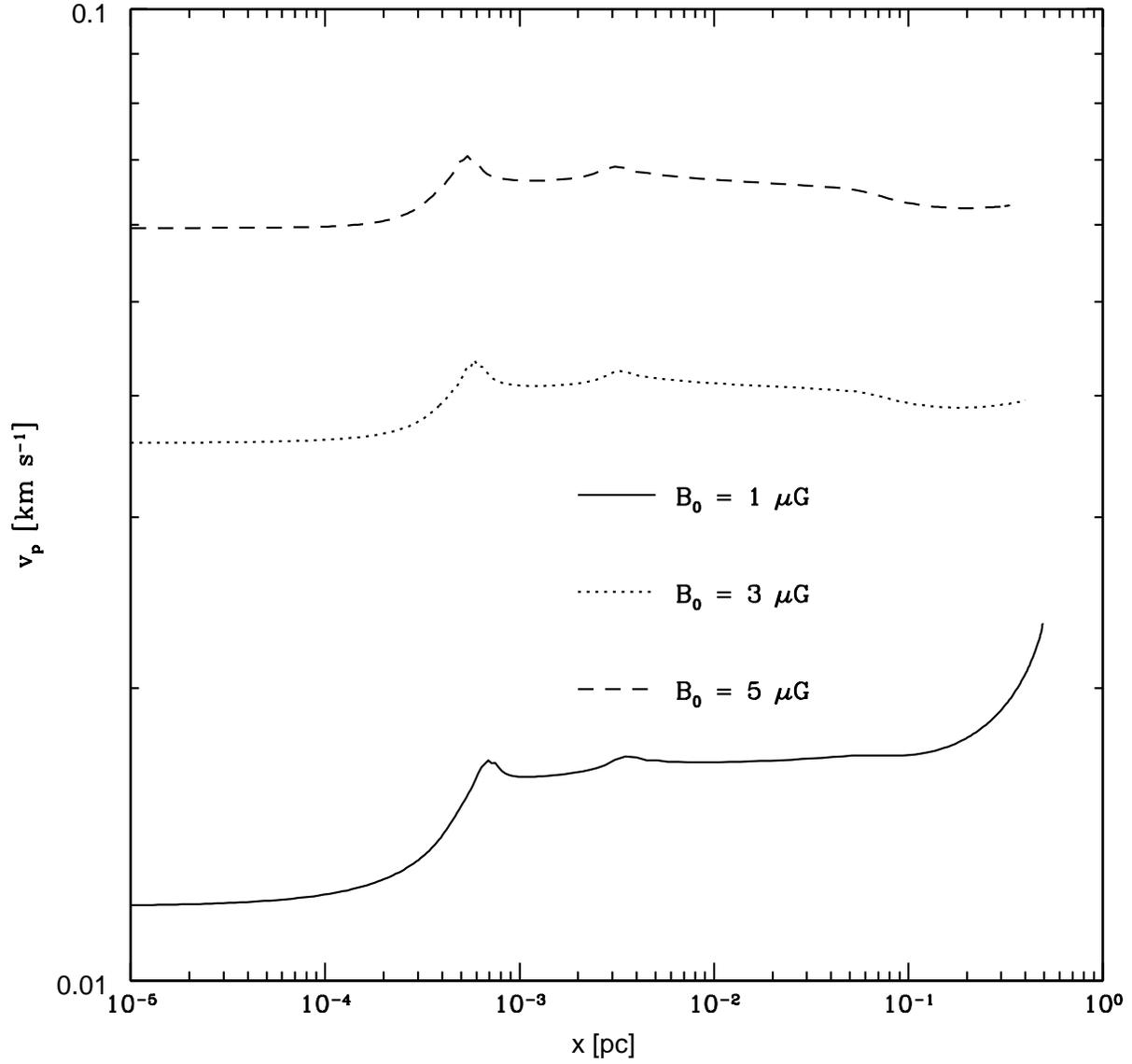}
\caption{\small Plasma velocity profiles of fronts having an initial density of $n = 106.08$ cm$^{-3}$ at 
various magnetic field strengths for fixed $|dB/dx|_{x_1} = 308.6 \hspace{0.1cm} 
\mu$G pc$^{-1}$.
}
\label{Bdriftfig}
\end{figure}

In Figure \ref{Bdriftfig} we plot the plasma velocity profiles of each of the front
models of Figure \ref{Bfig}. The size of the plasma velocity increases with 
increasing magnetic field strength, and the shape of the profile becomes
flatter. This is also due to the higher efficiency of ambipolar diffusion at
larger magnetic field strengths.

\begin{figure}
\epsscale{1.0}
\plotone{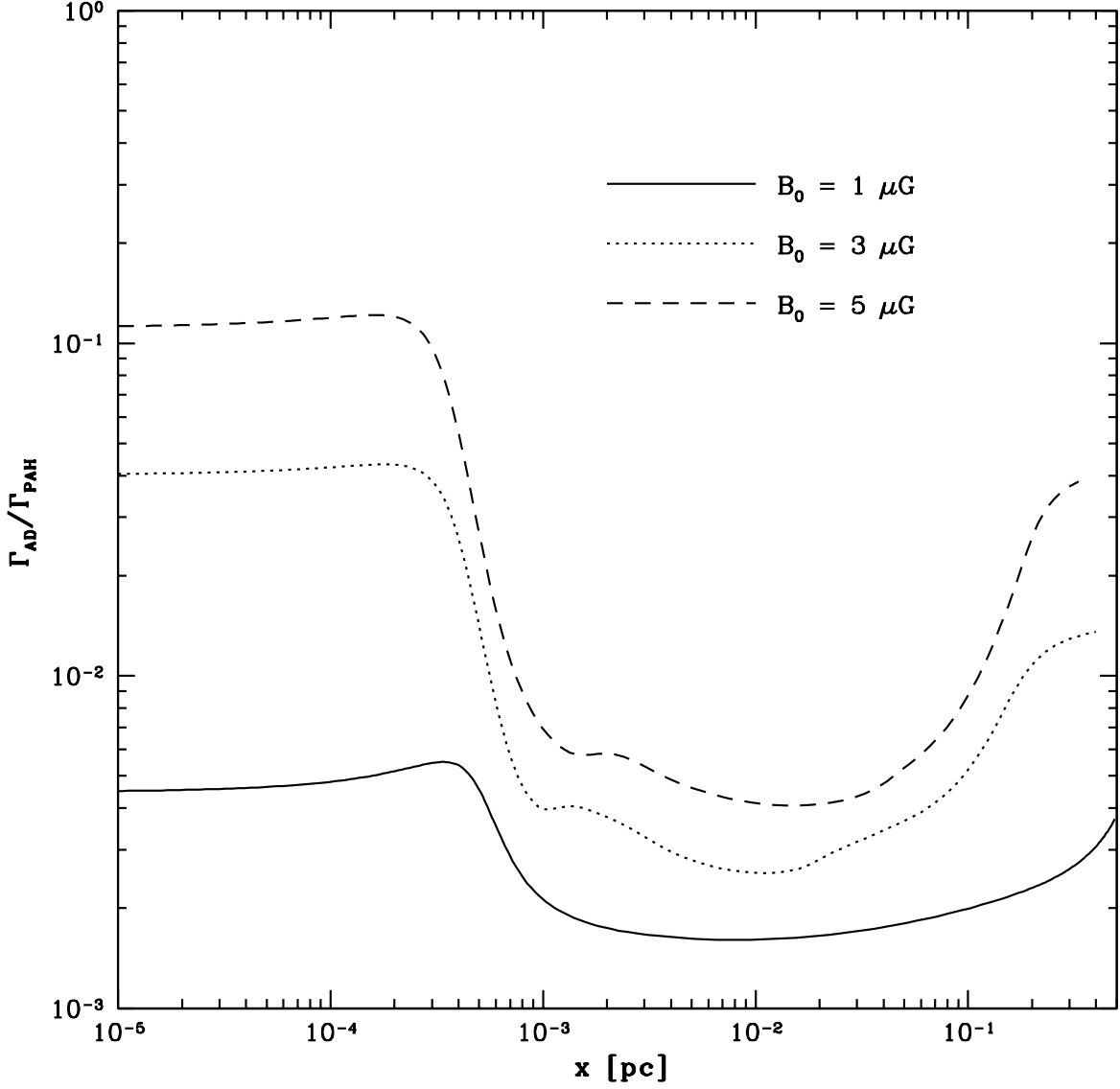}
\caption{\small Ambipolar drift heating rates normalized by photoelectric 
heating rate across fronts having an initial density of $n = 106.08$ cm$^{-3}$ 
and various initial magnetic field strengths for fixed $|dB/dx|_{x_1} = 308.6 
\hspace{0.1cm} \mu$G pc$^{-1}$. 
}
\label{Bheat}
\end{figure}

In Figure \ref{Bheat} we compare the 
photoelectric and ambipolar heating rates for the front models shown in Figure
\ref{Bfig}. For these particular cases the photoelectric heating rate is larger than 
the ambipolar heating rate for all field strengths, and the ambipolar heating 
rate increases with magnetic field strength.

\subsection{Flux-Freezing Approximation}
For completeness, we also present the flux-freezing approximation, in which the
behavior of the magnetic field is tied to the density such that $B/\rho$ is 
constant in 1-D. This result can be obtained by computing the total derivative 
of the quantity $B/\rho$ using the continuity and induction equations. To 
calculate the structure of a front in this approximation, we solve Equations 
(\ref{nrg2}) and (\ref{vev}) and everywhere replace $B$ by $\rho\cal{C}$, where 
$\cal{C}$ is a constant. Including the definition $z \equiv dT/dx$, we have a 
system of three ODEs, which we solve using our shooting method, with the mass 
flux, $j$, the parameter to be adjusted. We impose the boundary conditions given 
by equations (\ref{Tbc}) and (\ref{dTbc}), with no need for a condition on the 
magnetic field strength since its behavior is governed by that of the density.

\begin{figure}
\epsscale{1.0}
\plotone{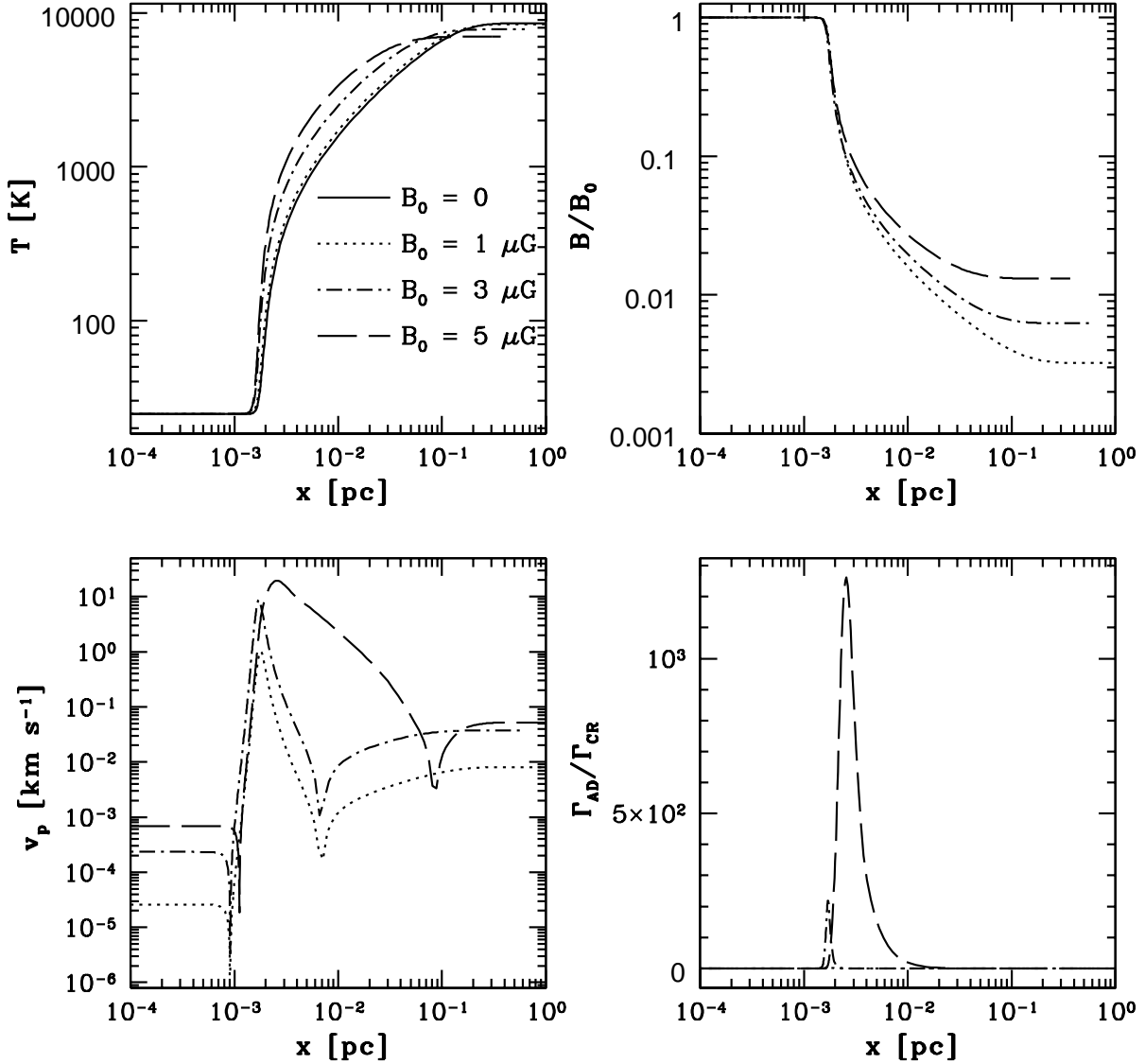}
\caption{\small Profiles of fronts having an initial density of $n = 106.08$
cm$^{-3}$, hence $p_{\rm thermal}/k_B = 2612$ K cm$^{-3}$, calculated in the 
flux-freezing approximation (without ambipolar drift heating) at various 
magnetic field strengths. The top panels show the temperature and magnetic field 
strength profiles. Although not shown here, the density profile has the same 
shape as the field strength profile, as dictated by flux-freezing. The lower 
panels show the plasma velocity and the ratio of the 
ambipolar heating rate to the photoelectric heating rate. The solid line shows the 
hydrodynamic result so only appears in the upper left panel.
\label{ff}}
\end{figure}

Figure \ref{ff} shows solutions for an initial density of $n = 106.08$ cm$^{-3}$ 
at various initial magnetic field strengths. As the field strength is increased 
the front becomes thinner and the transition reaches a progressively lower 
temperature, higher density final state at the right-hand boundary. Both the 
density and magnetic field strength span more than two orders of magnitude from 
one phase to the other. While such a range of densities is routinely observed in 
the neutral ISM, such widely varying magnetic field strengths are not (e.g. 
Troland \& Heiles 1986), and this provides the first indication that the 
flux-freezing approximation is not suitable for our problem.  

We go on to use these results to calculate ambipolar drift velocities, using 
Equation (\ref{drift}), and heating rates, given by Equation (\ref{adheat}), 
throughout the front. These are plotted in the lower two panels of Figure
\ref{ff}. For the most extreme case shown, a saturated front with an initial 
density of $106.08$ cm$^{-3}$ and an initial magnetic field strength of $5 
\hspace{0.1cm} \mu G$, we obtain a maximum drift velocity of $19.4$ km s$^{-1}$ 
and a maximum heating rate of $1.9 \times 10^{-21}$ ergs \hspace{0.1cm} s$^{-1}$ 
cm$^{-3}$, three orders of magnitude greater than the photoelectric heating rate. Although 
the equation for drift velocity breaks down for cases in which it is supersonic 
we may still employ it to show that if the flux-freezing approximation held, the 
drift velocities and heating rates would be enormous. Such an outcome is not 
self-consistent with the rest of the model, and allows us to argue that the 
solutions must be closer to what we have already presented, with the magnetic 
field strength almost constant over the extent of the front for cases in which 
ambipolar drift heating does not dominate. We thus suggest that by the time 
steady-state fronts are established in the neutral ISM the flux-freezing 
approximation does not apply. 

\section{DISCUSSION}
We have shown the magnetic field strength profiles of fronts having ion-neutral 
drift heating rates much smaller than the photoelectric heating rate to be almost 
flat. In this section we argue that it is the thin extent of these fronts that 
mediates the leakage of the magnetic field by ambipolar diffusion. We begin by 
using our results to justify our steady-state and ionization equilibrium 
assumptions. The minimum flow time through a front is of order $\tau_{flow} \sim 
0.01$ km s$^{-1} / 0.1$ pc $\sim 10$ Myr (refer to Figure \ref{prof1}). This 
should be compared to the ion-neutral collision time, given by $\tau_{in} \sim 
(\rho_n \gamma_{AD})^{-1} \sim 15.8 / n_n$ yr. Thus, we have 
$\tau_{in}/\tau_{flow} \ll 1$ so are safe in our steady-state formulation of
ambipolar diffusion. The 
assumption of ionization equilibrium is scrutinized by comparing $\tau_{flow}$ 
to the recombination time for hydrogen, given by $\tau_{rec} \sim 
1/\alpha^{(2)} n$, where $\alpha^{(2)} \sim 2.06 \times 10^{-11} T^{-1/2}$ 
cm$^{3}$ s$^{-1}$ \citep{Spitzer}. For a front with an initial density of 
$106.08$ cm$^{-3}$ we calculate a recombination time of $\sim 70$ yr on the cold 
side, and on the warm side we obtain $\sim 5000$ yr. For all our other front 
models we also find $\tau_{rec}/\tau_{flow} \ll 1$, so for this work our simple 
single-fluid treatment of ambipolar diffusion will suffice.  

We now present a diffusive description of fronts in which we compare the thermal 
and ambipolar diffusivities. Taking $U = n k_B T$ to be the energy density, we 
write the thermal timescale as $\tau_{th} = U/\rho \cal{L} \rm$, and the thermal 
diffusivity as $\lambda_{th} = \kappa T/U$, such that the characteristic length 
scale of the problem, the Field length, is given by $l_F = \sqrt{\lambda_{th} 
\tau_{th}}$ (Field 1965; Begelman \& McKee 1990). Hence, the thermal timescale 
and flow velocity may be written in terms of the thermal diffusivity, such that 
$\tau_{th} \sim l_F^2/\lambda_{th}$ and $v_{th} \sim \lambda_{th}/l_F$, 
respectively. In the magnetic field case, the field is redistributed diffusively, 
with ambipolar diffusivity, $\lambda_{AD} = v_A^2 \tau_{ni}$, where $\tau_{ni}$ 
is the neutral-ion collision time, approximated by $\tau_{ni} \sim 1.58 \times 
10^3/n_i$ yr \citep{PZN00}. Comparing the thermal and ambipolar diffusivities we
obtain:
\begin{equation}
\frac{\lambda_{\rm th}}{\lambda_{\rm AD}} = \frac{\kappa} {n k_B v_A^2 \tau_{ni}}
\sim 10^{-2} \frac{n_i T^{1/2}}{B_{\mu}^2} ,
\label{diffuse}
\end{equation}
where $B_{\mu}$ is the field strength in units of $\mu$G. We compute Equation
(\ref{diffuse}) at both boundaries of our front models and for all cases obtain
$\tau_{AD}/\tau_{th} \ll 1$. For example, for a front with an initial density 
and magnetic field strength of $106.08$ cm$^{-3}$ and $5 \hspace{0.1cm} \mu$G,
respectively, and $|dB/dx|_{x_1} = 308.6 \hspace{0.1cm} \mu$G pc$^{-1}$, we 
obtain $\tau_{AD}/\tau_{th} \sim 4.9 \times 10^{-5}$ on the cold side, and 
$\tau_{AD}/\tau_{th} \sim 4.0 \times 10^{-4}$ on the warm side. This means the 
drift time is always much smaller than the time to flow through the front, 
suggesting that the field has time to become close to uniform \footnote[1]{Note 
that the value of $|dB/dx|_{x_1}$ enters into this estimate only insofar as it 
affects the equilibrium temperature and front structure.}.

\begin{figure}
\epsscale{1.0}
\plotone{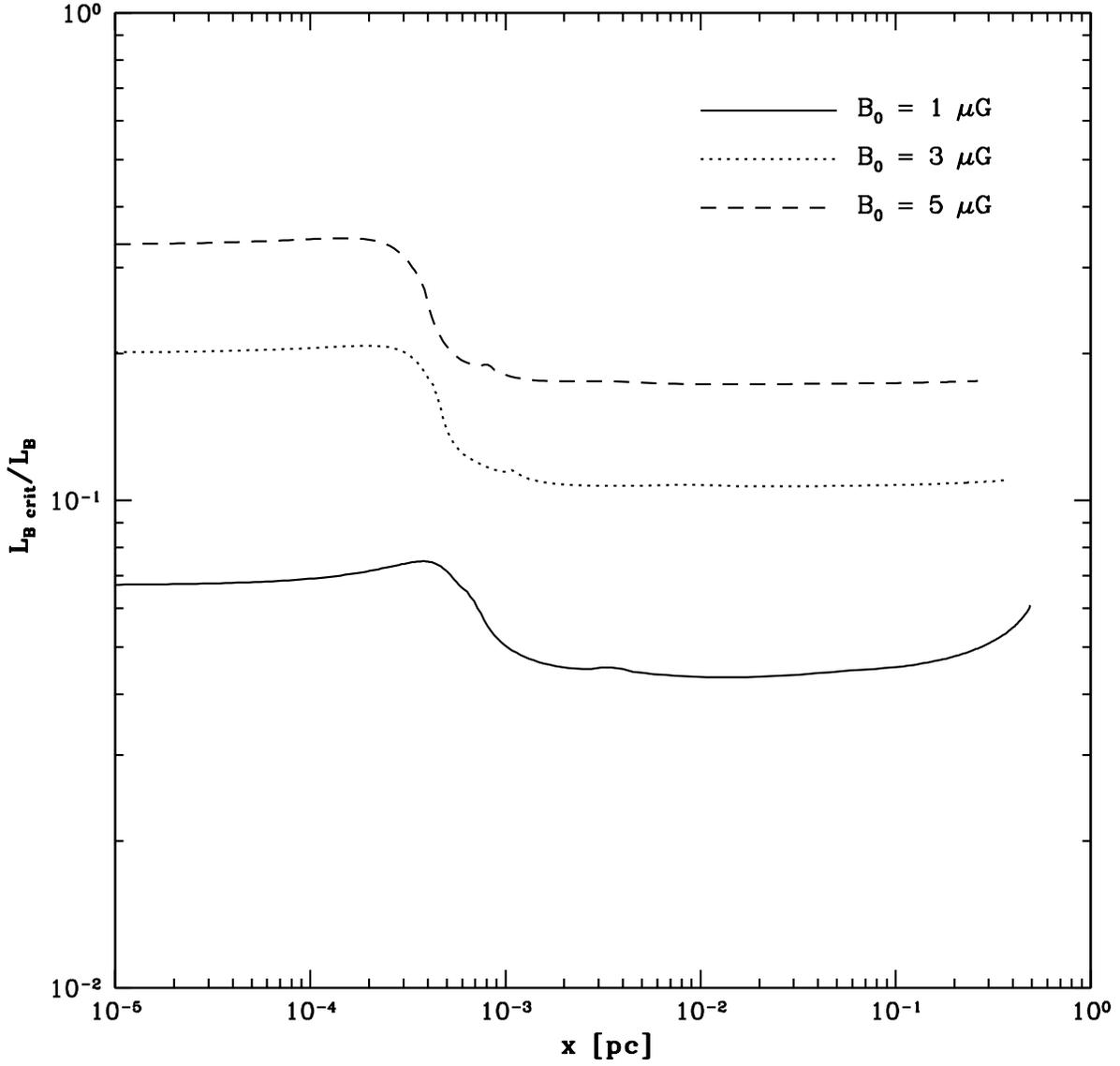}
\caption{\small Ratio of the critical magnetic length scale to the magnetic
length scale for fronts having an initial density of $n \sim 106.08$ cm$^{-3}$ 
and various initial magnetic field strengths, with $|dB/dx|_{x_1} = 308.6 
\hspace{0.1cm} \mu$G pc$^{-1}$. Ambipolar drift heating becomes important in 
determining the structure of the front if $L_B < L_{B crit}$. 
\label{lb}}
\end{figure}

Our results show that increasing the ambipolar heating rate changes the 
structure of our front solutions. By balancing the ambipolar and photoelectric 
heating rates, Equations (\ref{gammapah}) and (\ref{adheat}), and approximating 
the magnetic field strength gradient as $B_0/L_{B crit}$, we can estimate the 
critical length scale at which the magnetic field becomes important in 
determining structure:
\begin{equation}
L_{B crit}=\Bigg(\frac{\lambda_{\rm AD} B_0^2}{4 \pi n \Gamma_{\rm PAH}}\Bigg)^{1/2} .
\end{equation}
The magnetic length scale is given by $L_B \sim B/|\nabla B|$, so if $L_B > 
L_{B crit}$ the effect of the field on the structure of a front is small. We 
compare $L_B$ and $L_{B crit}$ in Figure \ref{lb} for a front with an initial 
density of $n \sim 106.08$ cm$^{-3}$ and initial field strengths of $B = 1$, 
$3$, and $5 \hspace{0.1cm} \mu$G, with an initial field strength gradient of 
$|dB/dx|_{x_1} = 308.6 \hspace{0.1cm} \mu$G pc$^{-1}$. For the $5 \hspace{0.1cm} 
\mu$G case we obtain $L_B \sim 1.6 \times 10^{-2}$ pc and $L_{B crit} \sim 5.4 
\times 10^{-3}$ pc on the cold side, and on the warm side we find $L_B \sim 
16.3$ pc and $L_{B crit} \sim 2.9$ pc. The magnetic length scale is larger than 
the critical scale throughout the front, thus ambipolar drift heating does not 
have a dramatic effect on the structure of a front. 

Previous dynamical studies have claimed that ion-neutral drift is not a 
sufficiently fast diffusion process for transporting magnetic energy, and 
instead invoked turbulent ambipolar drift \citep{HZ04} or turbulent magnetic
reconnection \citep{SL10} to explain the $B-\rho$ relation. However, these
studies were on larger scales than the fronts considered here. Our results 
suggest that for this simple scenario in which the phase transitions are thin, 
ambipolar diffusion alone is a sufficient mechanism for redistributing the 
magnetic field energy, without the need for turbulence. Our work directly
complements a study of the thermal instability as a formation mechanism for 
diffuse \ion{H}{1} clouds \citep{IIK07}. In that work it was shown that ambipolar diffusion
is a necessary and sufficient ingredient for the formation of a two-phase medium. Once that
medium is established the methods discussed in this paper may be applied to calculate its 
structure. 

\section{SUMMARY AND CONCLUSIONS}
In this work we have investigated the effect of magnetic fields on two-phase
structure in the neutral ISM. We have presented a numerical method for 
calculating the 1-D structure of fronts separating the cold neutral medium from 
the warm neutral medium, including the effects of ambipolar diffusion. We showed 
that the pressure range over which two-phase structure is permitted becomes 
larger, by as much as a factor of two, due to the contribution of ambipolar drift 
heating, 
with both the minimum and maximum pressures increasing from their hydrodynamic 
values. We find our magnetized front profiles to be very similar to the 
hydrodynamic solutions, and, in cases where photoelectric heating dominates 
ambipolar drift heating, to have close to flat magnetic field strength profiles. 
We also showed that the flux-freezing assumption yields unphysically large drift 
velocities and frictional heating rates. Our method is generic and, by including 
the appropriate physics, may be extended to other astrophysical multi-phase 
systems. 

Although the 1-D picture discussed in this work is fairly simple, if the
magnetic field strength and density were related we would have expected to see a 
correlation. Our results are consistent with the observational evidence that 
there is no relationship between magnetic field strength and density in 
interstellar atomic gas, which suggests that ambipolar diffusion is an efficient 
transport mechanism in the neutral ISM. The effect of ambipolar diffusion on the 
stability properties of thermal fronts will be the subject of forthcoming 
publications.

\acknowledgments
We acknowledge support from NASA ATP Grant NNGO5IGO9G and NSF Grant PHY0821899.
Our work has benefited from useful discussions with J. E. Everett, F. Heitsch, 
K. M. Hess, A. S. Hill, N. A. Murphy, L. M. Nigra, and R. H. D. Townsend. We
thank the referee for useful comments that enabled us to improve our manuscript.



\clearpage

\clearpage

\clearpage

\clearpage

\clearpage

\clearpage

\clearpage

\clearpage

\clearpage

\clearpage

\clearpage

\clearpage

\begin{deluxetable}{ccccccc}
\centering
\setlength{\tabcolsep}{0.040in}
\tabletypesize{\small}
\tablecaption{Properties of cold and warm phases connected by 
fronts\tablenotemark{1} having an initial density of 
$n_{\rm CNM} = 106.08$ cm$^{-3}$ and an initial magnetic field strength of 
$B_{\rm CNM} = 3 \hspace{0.1cm} \mu$G.}
\tablewidth{0pt}
\tablehead{
  \colhead{$|dB/dx|_{x_1}$ ($\mu$G pc$^{-1}$)} & \colhead{Front Type} &
  \colhead{$T_{\rm CNM}$ (K)} &
  \colhead{$n_{\rm WNM}$ (cm$^{-3}$)} & \colhead{$T_{\rm WNM}$ (K)} &
  \colhead{$B_{\rm WNM}$ ($\mu$G)} & \colhead{Thickness (pc)}
}
\startdata
{$0.31$} & Static & $24.63$ & $0.31$ & $8580$ & $3.000$ & $0.94$ \\
{$617.2$} & Condensation & $25.65$ & $0.45$ & $8210$ & $2.386$ & $0.28$ \\
{$1157.3$} & Condensation & $27.97$ & $0.67$ & $7818$ & $1.053$ & $0.12$ \\
{$1219.0$} & Evaporation & $28.30$ & $0.72$ & $7743$ & $0.037$ & $0.08$ \\
\tableline
\enddata  
\tablenotetext{1}{The profiles of the connecting fronts are presented in Figure
\ref{prof1}.}
\end{deluxetable}

\end{document}